\documentclass[sigplan,10pt]{acmart}
\renewcommand\footnotetextcopyrightpermission[1]{}

\usepackage{amsmath}
\usepackage{array}
\usepackage{amsfonts}
\usepackage{graphicx} 
\usepackage{amsthm}
\usepackage{enumitem}
\usepackage{booktabs}
\usepackage{xcolor}

\usepackage{algorithm}
\usepackage{algpseudocode}

 \AtBeginDocument{%
   }

\setcopyright{acmlicensed}
\copyrightyear{2026}
\acmYear{2025}
\acmDOI{XXXXXXX.XXXXXXX}

\acmConference{}{}{}

\begin{document}
\pagestyle{plain}

\title{Splitting Prompt Prefill from Response
Replay for Context-Parallel Long-Context LLM Post-Training}

\author{Yubing Bao}
\email{ybbao23@m.fudan.edu.cn}
\orcid{0009-0007-6753-6415}
\affiliation{%
  \institution{Fudan University}
  \city{Shanghai}
  \country{China}
}

\author{Zhihui Lu}
\email{lzh@fudan.edu.cn}
\affiliation{%
  \institution{Fudan University}
  \city{Shanghai}
  \country{China}
}

\author{Qiang Duan}
\email{qduan@psu.edu}
\affiliation{%
  \institution{The Pennsylvania State University}
  \city{State College}
  \country{USA}
}

\author{Yuedong Xu}
\email{ydxu@fudan.edu.cn}
\affiliation{%
  \institution{Fudan University}
  \city{Shanghai}
  \country{China}
}

\author{Sen Liu}
\email{senliu@fudan.edu.cn}
\affiliation{%
  \institution{Fudan University}
  \city{Shanghai}
  \country{China}
}

\author{Pan Zhou}
\email{panzhou@smu.edu.sg}
\affiliation{%
  \institution{Singapore Management University}
  \city{Singapore}
  \country{Singapore}
}


\renewcommand{\shortauthors}{Bao et al.}
\begin{abstract}
Training long-context LLM policies with RL requires re-evaluating groups of sampled responses under the updated policy, an update-stage attention workload that differs sharply from pre-training: each group shares one long prompt that fans out into multiple response branches. Standard context parallelism (CP) flattens each prompt--response pair into a linear sequence, so the same prompt key--value (KV) states are recomputed---or repeatedly rotated through the network---once per response branch. We present \textbf{AugTree}, a CP execution scheme built around this replay stage. AugTree separates the replay into two phases: a prompt-prefill phase that computes the shared prompt KV state once, and a response-replay phase that schedules the independent response branches over a bounded set of replay lanes. The replay phase instantiates two communication semantics, chosen by a lightweight online planner that enumerates CP degrees, schedules, and placements before GPU dispatch: rotating KV shards within response-local lanes when responses dominate, and moving response queries to stationary prompt-KV owners with a partial-softmax reduction when prompts dominate. The shared prompt state remains fully differentiable---response losses backpropagate into it and the accumulated prompt gradients propagate through the original prefill graph---so AugTree preserves exact training semantics rather than performing detached, inference-style KV caching. On four real post-training workloads and up to 64 accelerators, AugTree improves average training-stage step time by 1.18$\times$ over dynamic CP (up to 2.23$\times$), 2.63$\times$ over a Megatron ring CP baseline with prompt reuse, and 7.08$\times$ over the baseline without reuse.
\end{abstract}



\maketitle

\section{Introduction}
		\begin{figure}[t]
		\centering
		\includegraphics[width=0.95\linewidth]{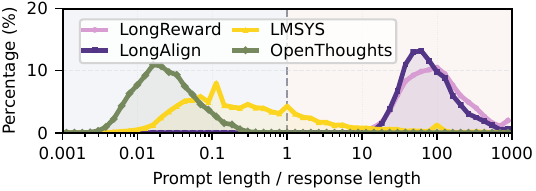}
		\vspace{-0.5em}
		\caption{\textbf{Prompt--response ratios vary widely.} In real post-training datasets, prompts can be much shorter or much longer than responses.}
		\label{fig:motivation-ratio-dist}
				\vspace{0em}
	\end{figure}

	Reinforcement Learning (RL) post-training has become a key mechanism for further improving Large Language Models (LLMs), with recent systems demonstrating substantial gains in mathematics, coding, agentic tasks and many others~\cite{guo2025deepseek,shao2024deepseekmath,yang2025qwen3,comanici2025gemini,openai2025introducing, deepseek2026deepseek, silvestre2025systems}. A standard RL post-training iteration involves three distinct stages: \textit{rollout} for sampling responses, \textit{reward} for scoring prompt--response trajectories, and \textit{training} for updating the model. Meantime, a growing line of rollout-side optimizations has substantially reduced rollout time: TRACE~\cite{zou2026trace} profiles current RL pipelines and finds rollout accounting for only 28.6\%/21.2\% of wall-clock time (policy update: 66.5\%/75.4\%), and DORA~\cite{hu2026dora} shrinks the rollout fraction from 65\% before optimization to 12\% after; Seer~\cite{qin2025seer} and NeMo-RL~\cite{silvestre2025systems} provide consistent evidence. As rollout becomes faster, update-side work becomes more important. AugTree targets precisely this update stage.
	
	This training bottleneck is further amplified by rapidly expanding LLM context windows. Recent models support very long contexts, e.g., 256K tokens in GPT-5.2~\cite{openai2025introducing}, 1M tokens in DeepSeek V4~\cite{deepseek2026deepseek} and 2M tokens in Gemini 2.5 Pro~\cite{comanici2025gemini}, substantially increasing training computation and memory cost. Context Parallelism (CP) addresses this pressure by partitioning long sequences across GPUs~\cite{liuringattention, jacobs2023deepspeed, gu2024loongtrain, jiang2025dcp}. Existing CP systems rely on Ring-style KV rotation~\cite{liuringattention, zhu2024ring}, Ulysses-style all-to-all shuffles~\cite{jacobs2023deepspeed}, hybrid designs~\cite{fang2024usp, gu2024loongtrain, nvidia2024transformer}, or dynamic optimization based on input sparsity~\cite{jiang2025dcp, wang2025flexsp, ge2025bytescale}. But they are designed for pre-training, where inputs are treated as independent sequences (only prompts, no responses), and thus fail to exploit the structured prompt--response groups in RL post-training, causing significant inefficiency.
	
	
	The key mismatch is that RL post-training data are not independent long sequences. Instead, they have three properties that shape how CP should compute, communicate, and plan. \textit{(1) Shared-prompt response groups.}
	For each prompt, RL post-training typically samples $G$ responses, e.g., $G=64$. Thus, each batch contains $G$ trajectories sharing a prompt but branching into different responses, forming a natural tree: one shared prefix with multiple response paths. The attention dependency is asymmetric: responses attend to the prompt, while the prompt need not attend to responses. \textit{(2) Highly variable prompt--response ratios.}
	As shown in Fig.~\ref{fig:motivation-ratio-dist}, our analysis of four real-world post-training datasets (LongAlign~\cite{bai2024longalign}, LongReward~\cite{zhang2025longreward}, LMSYS~\cite{zheng2024lmsys}, OpenThoughts~\cite{guha2025openthoughts}) shows that prompt--response ratios vary by up to a hundredfold. Long-document QA is often prompt-heavy, whereas creative writing is response-heavy. This variation determines which CP cost dominates: moving shared-prompt states or processing long response branches. \textit{(3) Online data generation.}
	Unlike pre-training data that can be packed or planned offline, post-training trajectories are generated online during rollout. Thus, sequence lengths, response groups, and prompt--response ratios are known only after generation, requiring CP to make placement and communication decisions at update time rather than relying on static offline planning.

When standard CP is applied to such data, it flattens each prompt--response trajectory into an independent sequence and loses the tree structure. The first two properties above then become two major sources of waste, while the third makes them hard to eliminate with fixed offline plans.
First, flattening destroys prompt sharing. Since the same prompt appears in all $G$ trajectories, standard CP may recompute prompt-side states for every trajectory, whereas the shared prompt should ideally be computed once. Thus, up to $(G{-}1)/G$ of prompt-side computation is redundant, corresponding to $87.5\%$ redundancy when $G{=}8$ and $98.4\%$ when $G{=}64$. This waste follows directly from shared-prompt grouping. Second, flattening causes unnecessary attention-state movement. Standard Ring-Attention-style CP usually moves KV blocks. For prompt-heavy groups, it may repeatedly move the same large prompt KV for different response branches, even though the prompt is shared. Here, moving response-side queries to stationary prompt KV can be cheaper. When responses dominate, moving KV may remain preferable, but communication should preserve response-path locality (i.e., keep each branch's tokens on the same CP rank) so one branch does not receive irrelevant KV from siblings. Thus, the best strategy depends on both the prompt--response ratio and the response-group tree.


	\begin{figure}[t]
		\centering
		\includegraphics[width=0.95\linewidth]{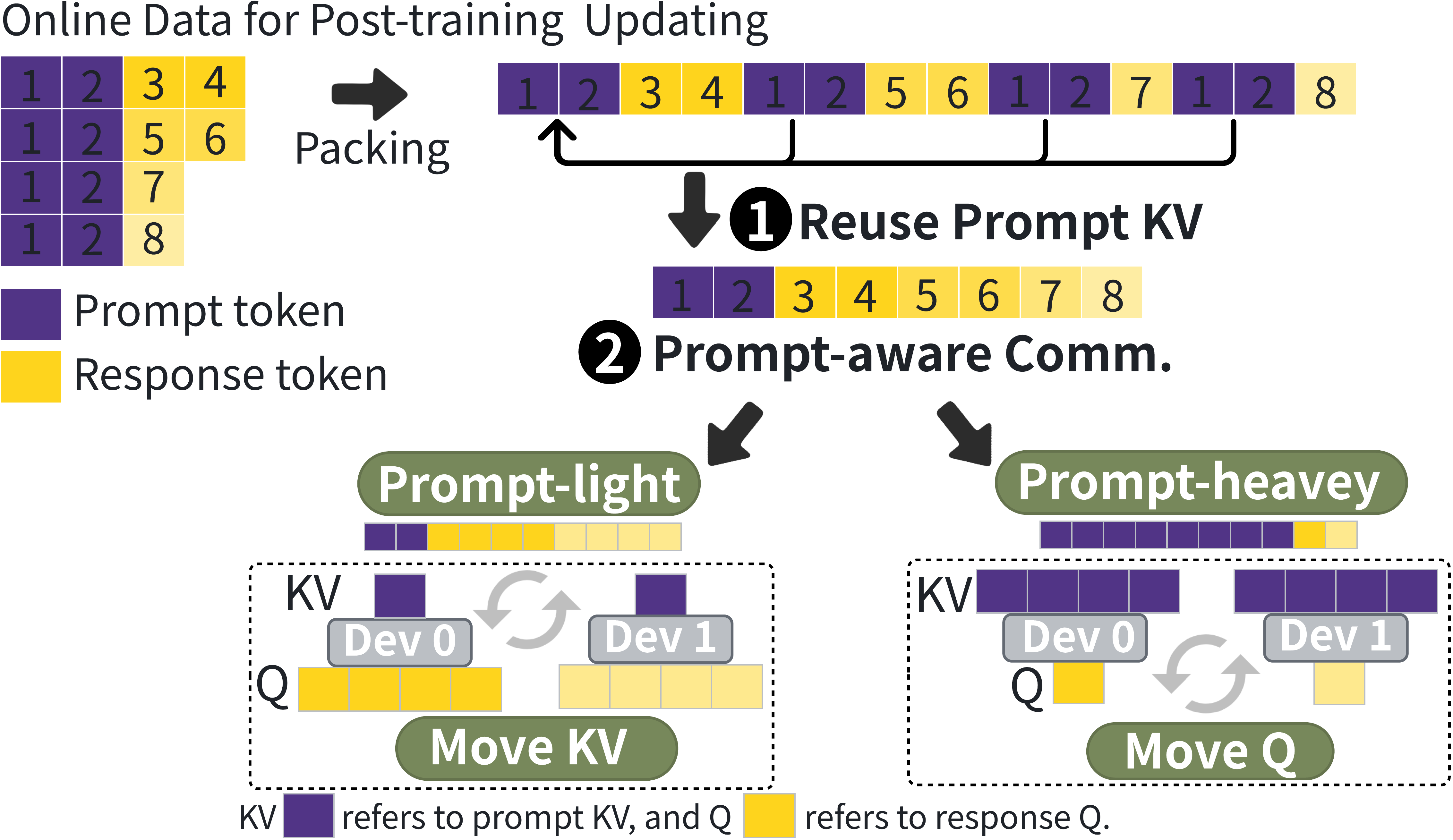}
						\vspace{-0.5em}
		\caption{\textbf{Two sources of AugTree's gain.} AugTree (1) computes the shared prompt KV once for all responses and (2) chooses whether attention communication should move KV or Q based on the prompt--response ratio.}
		\label{fig:intro}
					\vspace{0em}
	\end{figure}
    
To address these issues, we present \textbf{AugTree}, a prompt-aware context-parallel training method for long-context RL post-training. AugTree treats each response group as a prompt--response tree rather than $G$ independent sequences. This view leads to three components that directly match the three post-training properties above, as shown in Fig.~\ref{fig:intro}. \textit{\textbf{(1) Prompt KV reuse.}}
To exploit shared-prompt response groups and remove redundant prompt computation, AugTree computes the shared prompt KV once and reuses it across all responses from the same prompt. This applies to all prompt--response ratios and becomes more beneficial as prompt length or group size grows.
\textit{\textbf{(2) Prompt-aware communication.}}
To reduce unnecessary attention-state movement, AugTree chooses the CP communication strategy based on the prompt--response ratio. For 
\textit{prompt-dominant groups}, repeatedly moving large prompt KV is expensive. AugTree introduces \textit{MoveQ}, which keeps prompt KV and reduction state stationary, moving only response-side queries to prompt shards. A single global reduction then merges partial softmax states. For \textit{response-dominant groups}, moving KV can remain more efficient, but communication should preserve response-path locality whenever possible. AugTree therefore introduces \textit{TreeMoveKV}, a tree-aware move-KV primitive that avoids moving KV from irrelevant sibling responses. \textit{\textbf{(3) Online planning and execution.}}
To handle online data generation and varying prompt--response ratios, AugTree uses an asynchronous planner and a plan-driven execution engine: the planner selects an efficient CP plan after rollout, when trajectory lengths and group structure are known, and the execution engine applies it during training.
	Our contributions are:
	\begin{itemize}[leftmargin=*]
%
		\item To the best of our knowledge, AugTree is the first CP work to study grouped-response RL post-training as a shared-prompt workload. We identify its three key properties---shared prompts, variable prompt--response ratios, and online generation---and show that standard ring-style CP loses this structure, causing redundant prompt computation and unnecessary attention-state movement.

		\item We propose AugTree, a prompt-aware CP method for long-context post-training updates. AugTree models each response group as a prompt--response tree, computes the shared prompt once, reuses prompt KV across responses, and preserves response-path locality during CP placement.
		
		\item We introduce two communication primitives for different prompt--response regimes: \textit{MoveQ} for prompt-dominant groups and \textit{TreeMoveKV} for response-dominant groups. They avoid repeatedly moving large shared-prompt KV and irrelevant sibling-response KV, respectively.
		
		\item We implement AugTree with an asynchronous planner and plan-driven execution engine for post-training updates. On four real post-training workloads using up to 64 accelerators, AugTree improves average training-stage step time by 1.18$\times$ over DCP (up to 2.23$\times$), by 2.63$\times$ over MegaR (Megatron ring CP with prompt reuse) and 7.08$\times$ over MegaN (without prompt reuse).

	\end{itemize}

\section{Background and Motivation}
\label{sec:background}
We elaborate on the mismatch between RL post-training and existing
context-parallel (CP) training systems below.  

\subsection{Post-Training Replay as Prompt--Response Trees}
\label{sec:bg_replay_tree}

\noindent\textbf{Replay workflow.}
RL post-training refines an LLM after pre-training, improving capabilities such as
reasoning and tool use~\cite{comanici2025gemini,openai2025introducing,guo2025deepseek,yang2025qwen3}.
A standard iteration runs rollout (sampling $G$ responses $\{R_i\}_{i=1}^{G}$ for a prompt
$P$ from the current policy), reward (assigning each trajectory a score or advantage), and training
(replaying the scored trajectories and updating the policy). We use \emph{replay workload} for this
training-stage re-evaluation, not an off-policy replay buffer; AugTree targets exactly this stage.

\noindent\textbf{Replay tree.}
Replay data from one prompt is not a set of unrelated sequences. All $G$ responses share
the same prompt prefix $P$ and branch only after rollout, forming a prompt--response tree:
the prompt is the shared root, and the responses are independent sibling branches.
Existing training stacks often flatten this tree into $G$ independent sequences
$\{(P,R_1), (P,R_2), \ldots, (P,R_G)\}$.
Conventional CP training targets linear sequences. This flattening hides the shared prefix:
prompt-side states are recomputed or communicated repeatedly, as if each response owned a
private prompt copy. The waste is especially costly in long-context post-training, where $P$ may
contain long documents, retrieved contexts, conversation histories, or tool traces, so the length $|P|$ can far
exceed the  length $|R_i|$.

\begin{figure}[t]
	\centering
	\includegraphics[width=0.7\linewidth]{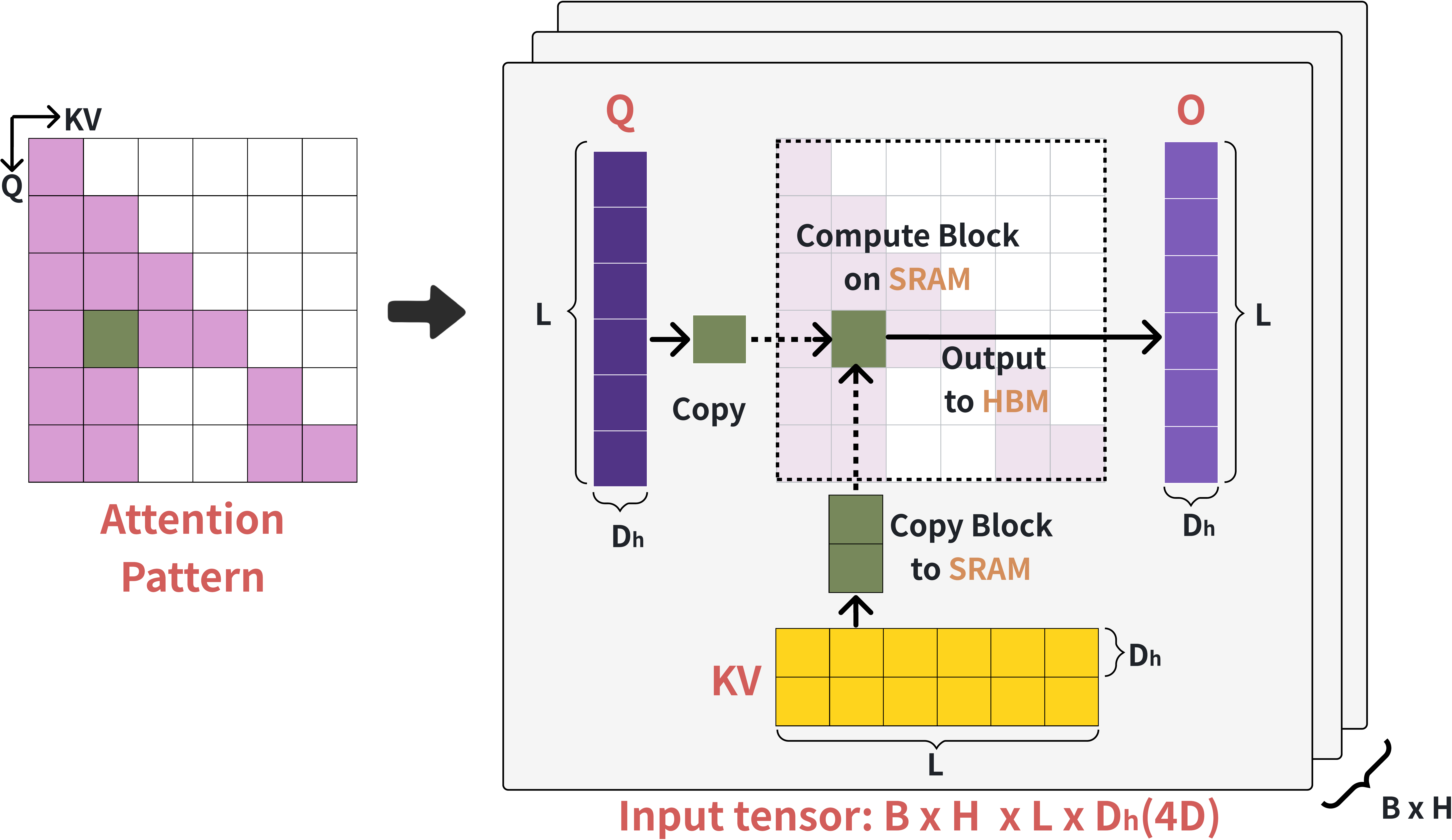}
	\vspace{-0.5em}
	\caption{\textbf{Attention as Q--KV block interactions.}
		Long-context CP schedules valid query--KV block interactions across devices.}
	\label{fig:attention}
	\vspace{0em}
\end{figure}

\noindent\textbf{Asymmetric KV dependency.}
For an autoregressive LLM, the likelihood of response $R_i$ factorizes as
$\pi_\theta(R_i|P)=\prod_t p_{i,t}$, where $p_{i,t}=\pi_\theta(R_{i,t}\,|\,P,R_{i,<t})$:
each response token attends to the full prompt and its own prefix, but not to sibling responses.
In the forward pass, response logits require the shared prompt KV states $KV_P$, but computing
$KV_P$ does not depend on any response tokens, so the prompt can be computed once and reused
across responses; in the backward pass, response-token losses propagate gradients through the
attention paths that read $KV_P$, so the prompt states must accumulate contributions from all
response branches with exact many-to-one gradient aggregation (Sec.~\ref{sec:prefill_once}).

\begin{figure}[t]
	\centering
	\includegraphics[width=0.9\linewidth]{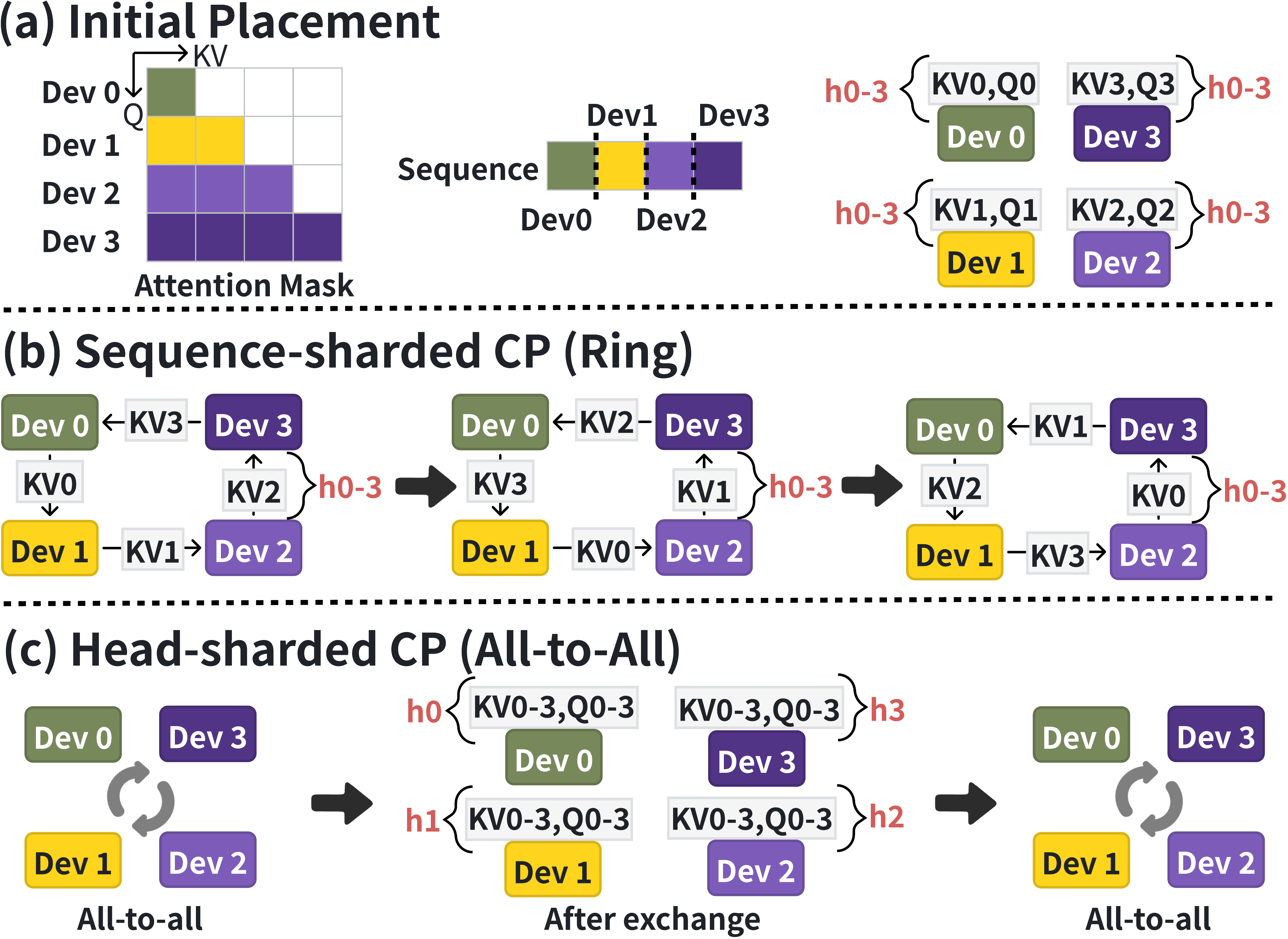}
	\vspace{-0.5em}
	\caption{\textbf{How existing CP moves attention tensors.}
		Sequence-sharded CP moves KV blocks to query owners, while head-sharded CP uses all-to-all
		tensor reshaping to make each head local.}
	\label{fig:ring}
	\vspace{0em}
\end{figure}
\subsection{Linear CP Assumptions and Replay Mismatch}
\label{sec:bg_cp_mismatch}

CP shards a long sequence across devices to reduce per-GPU KV memory; among transformer operators,
only attention requires coordinated access to remote keys and values, making communication a
first-order bottleneck.

\noindent\textbf{Attention as block interactions.}
For one attention head, let $\mathbf{Q},\mathbf{K},\mathbf{V}\in\mathbb{R}^{L\times D_h}$ denote the query,
key, and value tensors, where $L$ is the sequence length and $D_h$ is the per-head hidden dimension.
The attention output is
$\mathbf{O}=\mathrm{softmax}({\mathbf{Q}\mathbf{K}^{\top}}/{\sqrt{D_h}} + \mathbf{M})\,\mathbf{V}$,
where $\mathbf{O}\in\mathbb{R}^{L\times D_h}$ and the mask $\mathbf{M}$ marks invalid query--key positions.
Direct attention materializes an $L\times L$ score matrix, while FlashAttention-style kernels tile $\mathbf{Q}$ and $\mathbf{KV}$
blocks and maintain online softmax state in SRAM~\cite{dao2022flashattention}. As shown in
Fig.~\ref{fig:attention}, each valid tile computes one query--KV interaction and accumulates
partial softmax/output state. CP distributes these block-level interactions across devices.

\begin{figure}[t]
	\centering
	\includegraphics[width=0.9\linewidth]{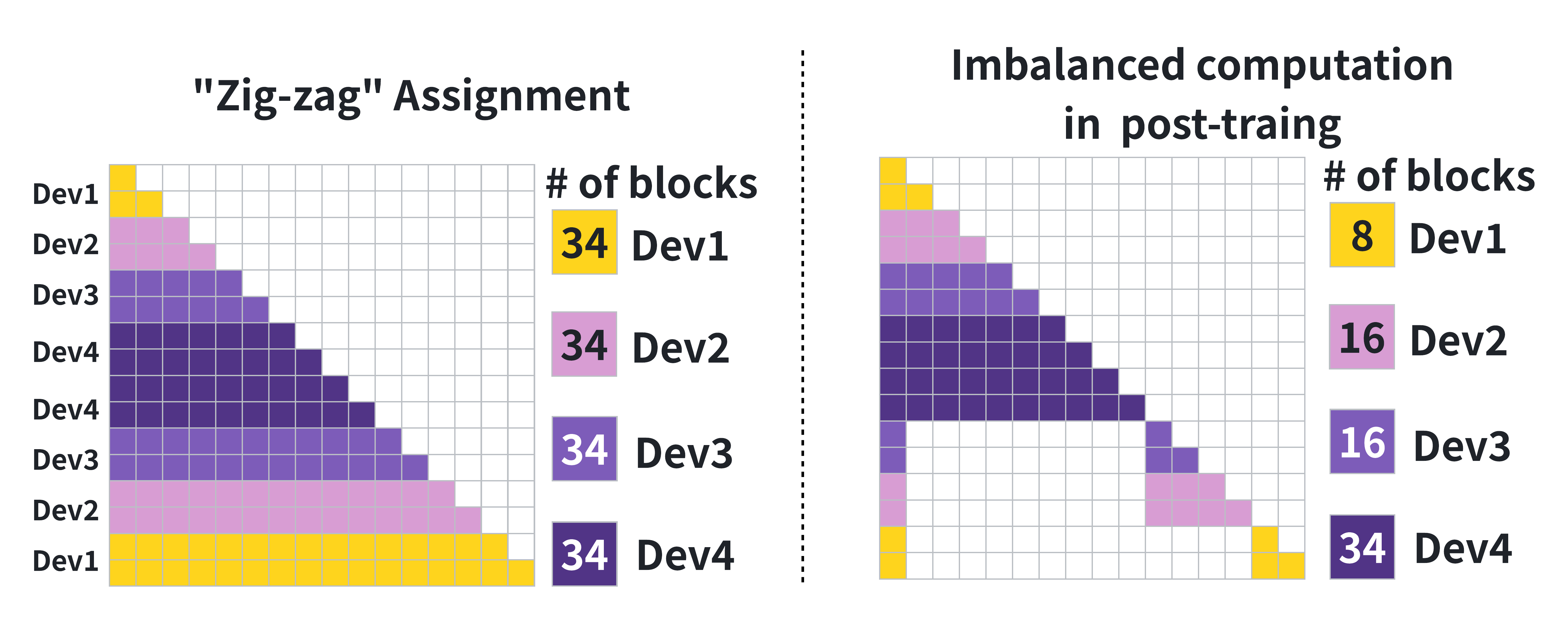}
	\vspace{-0.5em}
	\caption{\textbf{Zigzag load balancing.}
		Standard CP balances linear causal attention by pairing early and late sequence chunks. For
		tree-structured replay masks, however, the same placement can fragment branches and create load
		imbalance.}
	\label{fig:zigzag}
	\vspace{0em}
\end{figure}

\noindent\textbf{Linear CP assumptions.}
Existing CP systems often make two implicit assumptions. First, they treat the input as a
\emph{linear sequence}. Sequence-sharded CP partitions tokens across devices; each device owns a
subset of query tokens and obtains remote KV blocks through Ring-Attention-style KV rotation or
all-gather-like exchange~\cite{zhu2024ring,liuringattention}. The common semantic is that
\emph{KV is mobile} and moves to reconstruct the global attention context. Second, head-sharded
CP, such as Ulysses-style CP~\cite{jacobs2023deepspeed}, uses all-to-all communication to
redistribute tensors so that each device holds the full sequence but only a subset of heads. Modern
2D CP systems combine sequence and head sharding for scalability
~\cite{gu2024loongtrain,fang2024usp,jiang2025dcp,nvidia2024transformer,wang2025flexsp,ge2025bytescale,liang2026hexiseq}.
These designs differ in layout, but all assume attention is executed over packed linear sequences.

\noindent\textbf{Replay mismatch.}
Post-training replay breaks the linear-sequence assumption: replay attention is a shared prompt
column plus block-diagonal response branches, where a token on branch $i$ attends only to $P$ and
its own prefix $R_i^{<t}$, not to sibling branches $R_j$ ($j\neq i$) --- not a dense causal triangle
over a flattened sequence.
Linear CP hides this structure after packing. It may
repeatedly move the same large prompt KV across response branches, and may move sibling-response
KV that is unused by the current branch; zigzag placement, for example, can fragment response
paths and imbalance tree-structured masks (Fig.~\ref{fig:zigzag}). So post-training replay requires CP to preserve prompt sharing and
response-path locality, rather than only balancing a flattened sequence.

\subsection{Why Dynamic CP is Still Insufficient}
\label{sec:bg_dcp}
Dynamic CP systems, represented by DCP~\cite{jiang2025dcp}, improve over static CP by building a dependency graph and partitioning valid attention blocks across ranks. This adds placement flexibility and helps avoid some topology-unaware scheduling decisions. But  post-training replay requires more than dynamic block placement: it must also select  \emph{communication semantic} of distributed attention and exploit the deterministic prompt--response tree generated online after rollout.

\begin{figure}[t]
	\centering
	\includegraphics[width=0.9\linewidth]{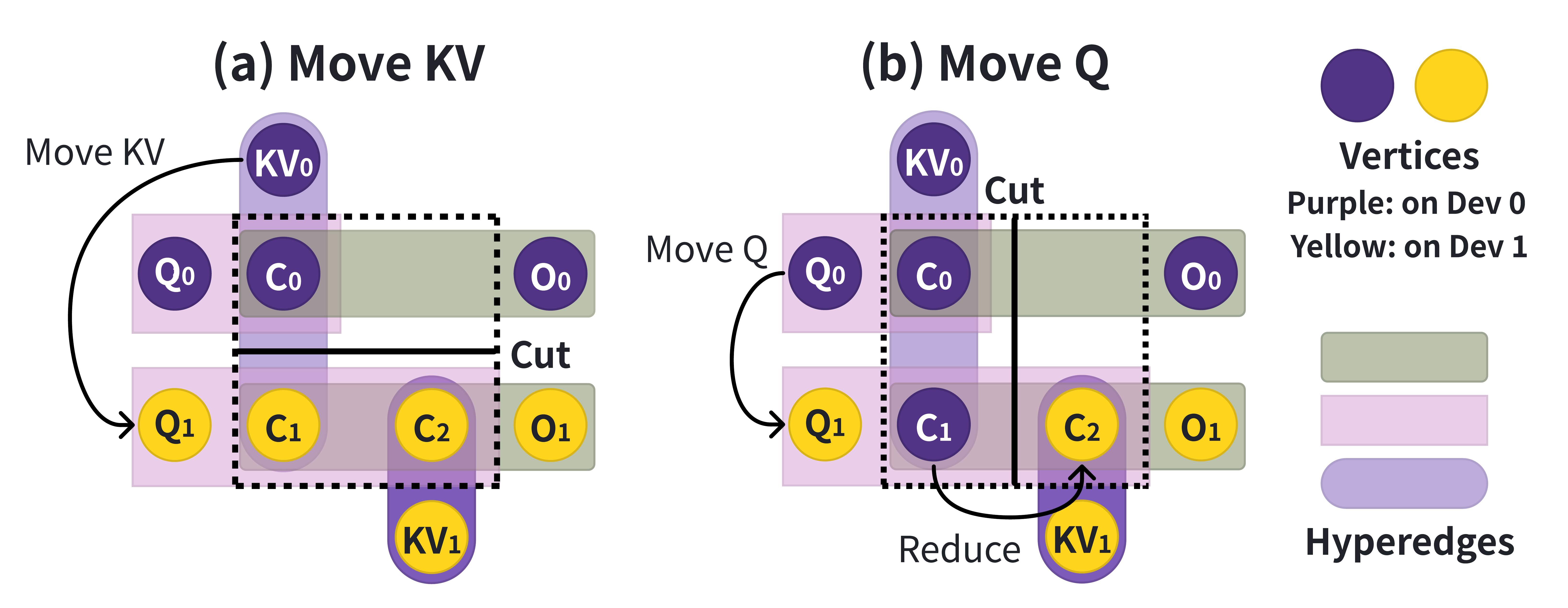}
	\vspace{-0.5em}
	\caption{\textbf{Moving KV and moving Q are different communication choices.}
		For the same valid attention tiles, a row-wise cut keeps query/output state local and moves KV,
		while a column-wise cut keeps KV local, moves Q, and reduces partial output states.}
	\label{fig:hypergraph}
	\vspace{0em}
\end{figure}

\noindent\textbf{Restricted communication semantics.}
For the same valid attention tiles, different executions preserve different tensors locally.
As Fig.~\ref{fig:hypergraph} illustrates, we view data blocks, tile computations, and intermediate
results as hypergraph vertices (one tensor block can feed multiple tiles) and their dependencies as
hyperedges. A row-wise cut keeps query/output state local and moves KV blocks --- the traditional
ring semantic; a column-wise cut keeps KV blocks local, moves query blocks, and reduces partial
softmax/output states. Both compute the same attention result, but their communication
costs differ. DCP optimizes placement under the traditional move-KV semantic, and therefore does
not search the move-Q alternative. This restriction can be costly in prompt-heavy replay, where a
large shared prompt KV is reused by many short responses and repeatedly moving it is inefficient.

\noindent\textbf{Missing online tree structure.}
DCP also treats each replay instance as a generic dependency graph. This misses two post-training
properties. First, replay data is generated online: sequence lengths, response groups, and
prompt--response ratios are known only after rollout. Second, the dependency graph is not arbitrary;
it has a deterministic tree structure with one shared prompt root and independent response branches.
Generic graph solving can add planning overhead and may fragment response paths, while a
tree-aware planner can directly exploit prompt sharing, path locality, and the prompt--response ratio
to select both placement and replay primitive.

\section{AugTree}
\label{sec:treecp}

\subsection{System Overview}
\label{sec:treecp_overview}

AugTree is plugged into the training stage of an RL post-training system, closing the gap identified in
Sec.~\ref{sec:background}: it reuses the shared prompt state, preserves response-path locality, and
chooses whether replay moves KV or Q before dispatch. Rollout and reward
engines first produce scored prompt--response groups, and AugTree treats each group as a replay tree
\(\mathcal{T}=(P,\{R_i\}_{i=1}^{G})\), where \(P\) is the shared prompt and \(R_i\) is the \(i\)-th
response branch. AugTree leaves rollout scheduling, reward computation, loss construction, gradient
synchronization, and optimizer steps unchanged; it only replaces the CP attention path
inside the trainer.

\begin{figure}[t]
	\centering
	\includegraphics[width=0.9\linewidth]{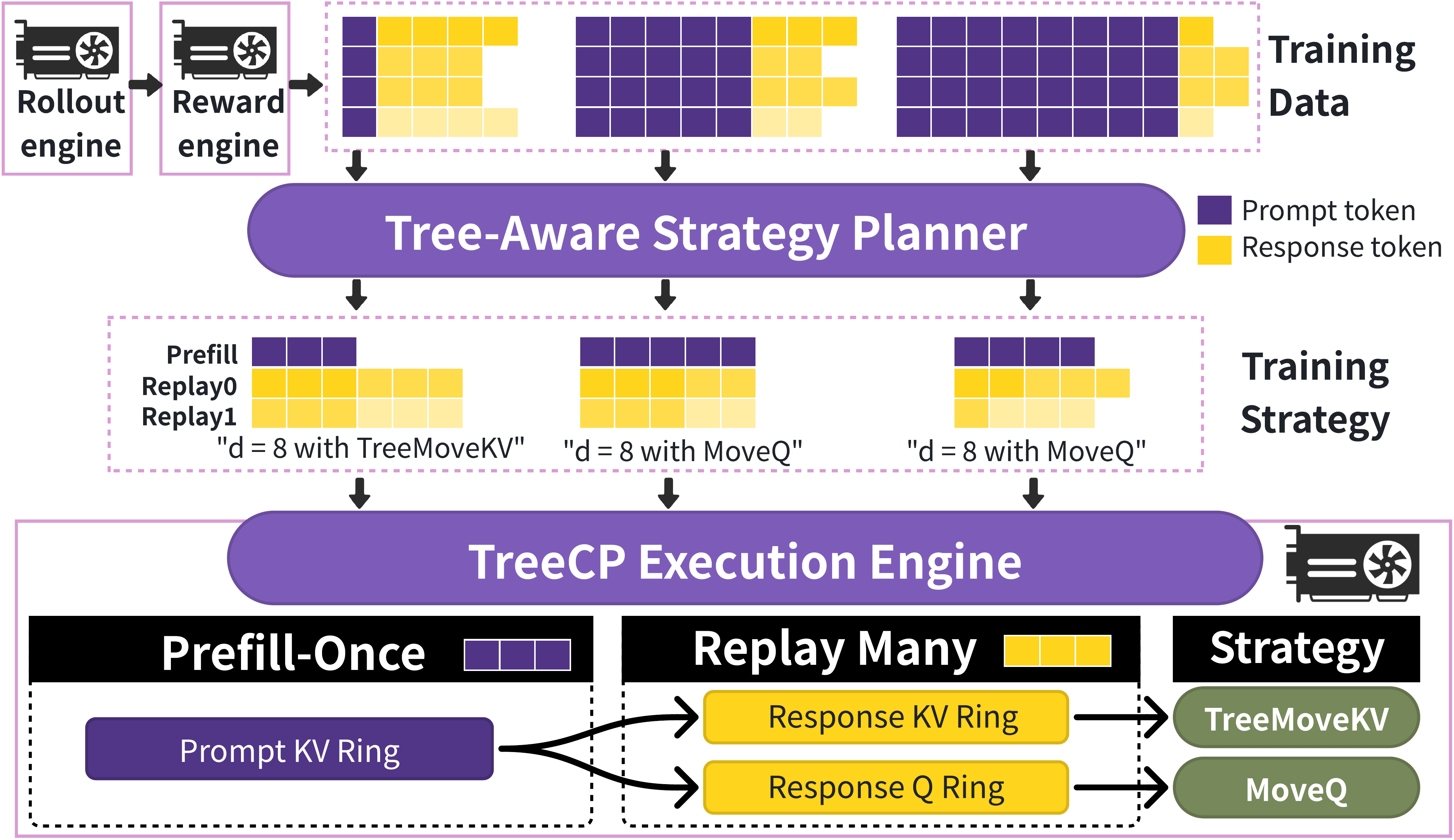}
	\caption{\textbf{AugTree planning and execution.}
		Rollout and reward produce scored prompt--response groups.
		AugTree plans over each replay tree, chooses the CP degree, response schedule, placement,
		and communication primitive, and executes the selected attention path inside the trainer.}
	\label{fig:overview}
	\vspace{0em}
\end{figure}

As shown in Fig.~\ref{fig:overview}, AugTree follows a plan-then-execute design. For each replay tree,
the planner chooses an execution plan
\(c=(d,\pi,\mathbf{r},X)\), where \(d\) is the CP degree,  \(\pi\) is the replay primitive, 
\(\mathbf{r}=(b,s)\) is the response-replay grid, and \(X\) maps response replay units to grid
locations. When \(b=0\), AugTree does not
separate prompt prefill from response replay; it falls back to ordinary ring CP and runs the prompt
and response tokens together. This fallback is useful for short groups, where the overhead of
tree-specific replay may outweigh its benefit. For split plans with \(b\ge1\), the planner chooses
\(\pi\in\{\textsc{TreeMoveKV},\textsc{MoveQ}\}\) and schedules the response branches over the replay
grid. 

Then the execution engine maps the selected plan to process groups, tensor movement, and attention
kernels. Section~\ref{sec:prefill_once} describes how AugTree establishes and
reuses the shared prompt state while preserving exact gradients;
Section~\ref{sec:primitive_selection} presents the two tree-aware replay
primitives; Section~\ref{sec:online_planner} defines the legal plan
space and the bounded online planner; and
Sec.~\ref{sec:execution_engine} explains how the selected plan is
realized inside an existing distributed trainer.

\subsection{Prefill-Once Prompt KV Reuse}
\label{sec:prefill_once}

For a split plan with \(b\ge1\), AugTree separates prompt establishment from response replay.
Given a replay tree \(\mathcal{T}=(P,\{R_i\}_{i=1}^{G})\), the valid KV context of the \(t\)-th token
on branch \(i\) is
\begin{equation}
	\label{eq:branch-kv-context}
	KV(R_{i,t}) = KV_P \cup KV(R_{i,<t}),
\end{equation}
with no \(KV(R_j)\) for \(j\neq i\). So AugTree  builds the prompt-side KV state once and
shares it across all response branches.

\noindent\textbf{Shared prompt prefill pass.}
When the planner selects a split plan $b\ge1$, AugTree first executes a single differentiable prompt-prefill pass, producing the sharded prompt state
$KV_P = \{(K_P^{(\ell)},V_P^{(\ell)})\}_{\ell=1}^{N_{\mathrm{layer}}}$ for each of the $N_{\mathrm{layer}}$ Transformer layers.
Since $KV_P$ is fixed during the subsequent response replay, all $G$ branches reuse this one state rather than materializing private prompt copies.

%

\noindent\textbf{Response replay.}
After prefill, AugTree replays responses with teacher forcing.
Each replay call feeds response tokens together with the cached prompt KV.
The response-side attention tiles are then executed using either \textsc{TreeMoveKV} or \textsc{MoveQ}, depending on the planner decision.
The first response token $R_{i,1}$ is scored by applying the language-model head to the final prompt hidden state from the prefill pass, and its loss backpropagates through that state into the prompt graph.

\noindent\textbf{Exact many-to-one backward.} 
Prompt reuse in training is not an inference-only KV cache.
Although losses are defined on response tokens, gradients still flow through attention paths that read $KV_P$.
Therefore, AugTree must accumulate prompt-state gradients from all response branches:
\begin{equation}
	\label{eq:prompt-gradient-accumulation}
	\nabla KV_P = \sum_{i=1}^{G} \nabla_{KV_P}\mathcal{L}_i,
\end{equation}
where $\mathcal{L}_i$ is the response-token loss on branch $i$.
When exact prompt gradients are enabled, the execution engine accumulates gradient seeds on the prompt-KV leaves and backpropagates them through the original prompt-prefill graph, preserving exact many-to-one gradient aggregation.

\noindent\textbf{Activation lifetime.}
The prompt computation graph remains logically live until all response
units in the replay tree have contributed their prompt-state gradients.
AugTree follows the checkpointing policy of the base trainer: prompt
activations required by that policy are either retained or recomputed
during the final prompt backward traversal. Once the aggregated prompt
gradient has been propagated, AugTree releases the prompt graph and its
shared-prefix state. This lifetime is scoped to one replay tree and
does not change the model parameters, optimizer state, or loss
definition.

\begin{figure*}[t]
	\centering
	\includegraphics[width=0.95\linewidth]{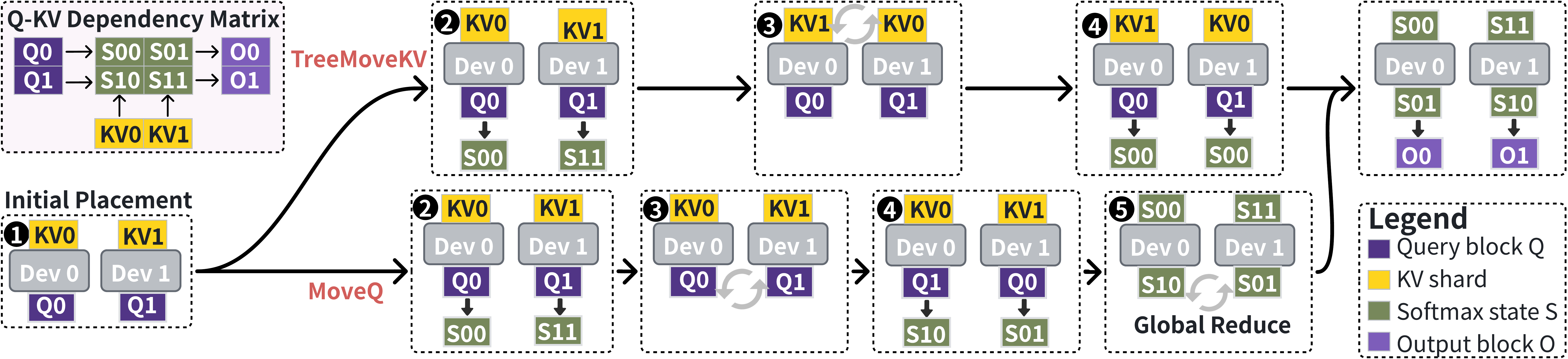}
	\caption{\textbf{TreeMoveKV and MoveQ on Q--KV dependencies.}
		Purple blocks are query blocks, yellow blocks are KV shards, green
		blocks are partial online-softmax states
		\(S=(m,\ell,O_{\mathrm{acc}})\), and light-purple blocks are final
		outputs. TreeMoveKV rotates KV shards while queries remain on their
		owners. MoveQ keeps KV shards stationary, sends queries to KV
		owners, and reduces the resulting partial softmax states.}
	\label{fig:moveq}
	\vspace{0em}
\end{figure*}

\subsection{Tree-Aware Replay Primitives}
\label{sec:primitive_selection}

After the shared prompt pass in Sec.~\ref{sec:prefill_once}, response replay must evaluate all valid
Q--KV tile interactions implied by Eq.~\ref{eq:branch-kv-context}.
For the same valid tile set, distributed attention can preserve
different tensors locally: it can keep queries and outputs local while
moving KV, or keep KV local while moving queries and reducing partial
outputs. AugTree instantiates these two communication semantics as
\textsc{TreeMoveKV} and \textsc{MoveQ}.

\noindent\textbf{Common tile semantics.}
As illustrated in Fig.~\ref{fig:moveq}, both \textsc{TreeMoveKV}  and \textsc{MoveQ} evaluate the same valid tiles in the Q--KV dependency matrix. Each interaction between a query block \(Q_i\) and a KV shard \(KV_j\) produces a partial online-softmax state \(S_{ij}\). Following standard online-softmax formulations~\cite{milakov2018online,dao2022flashattention}, all states along the dependency row of \(Q_i\) are associatively merged, e.g., \(S_{i0}\oplus S_{i1}\rightarrow O_i\), where \(\oplus\) denotes online-softmax state merging rather than ordinary addition. Thus, the two schemes produce identical attention outputs.


\noindent\textbf{TreeMoveKV: response-local move-KV.}
\textsc{TreeMoveKV} is used  when response replay dominates. It preserves the standard move-KV execution of ring-style CP~\cite{liu2024ringattention}: each rank retains its query and output blocks, while KV shards circulate across ranks and partial attention states are merged via standard online-softmax techniques~\cite{dao2022flashattention}. After one ring traversal, each query owner obtains its exact attention output.

The key innovation is \emph{tree-aware placement}. Rather than flattening the prompt and all sibling responses into a globally partitioned sequence, AugTree assigns each response path, or a small group of paths, to a replay lane of \(d_{\mathrm{lane}}\) ranks. Within each lane, only the shared prompt KV and the response KV required by its assigned paths are circulated.
Each lane prefills the full prompt KV sharded across its \(d_{\mathrm{lane}}\) members, so prefill cost repeats \(s=d/d_{\mathrm{lane}}\) times while per-rank memory stays at the single-lane share; this repetition is included in the planner cost model. At every ring stage, each rank evaluates the valid Q--KV tiles for its local queries and merges the resulting partial states. This lane-local execution prevents KV blocks from unrelated sibling responses from being communicated while retaining the low-overhead ring dataflow.

%

\noindent\textbf{MoveQ: KV-stationary replay.}
\textsc{MoveQ} targets prompt-heavy groups, where repeatedly circulating a large shared prompt KV state for many short responses would dominate communication. Instead, it keeps the canonical prompt KV shards stationary on their owning ranks and sends the smaller response-query blocks to the ranks holding their valid prompt or response KV shards. Each KV owner evaluates its local Q--KV tiles and produces a partial online-softmax state
\(
S=(m,\ell,O_{\mathrm{acc}})
\),
where \(m\) is the local row-wise maximum, \(\ell\) is the corresponding softmax normalizer, and \(O_{\mathrm{acc}}\) is the unnormalized output accumulator~\cite{milakov2018online,dao2022flashattention}. The partial states for the same query are then associatively merged, after which the exact normalized attention output is materialized on the original query owner.

The reduction operates on full partial states \((m,\ell,O_{\mathrm{acc}})\) rather than independently normalized outputs, preserving exact attention semantics; its custom autograd operator applies the reverse communication so gradients return to the original query and KV owners. Thus, \textsc{MoveQ} replaces repeated movement of the dominant prompt KV state with response-query communication and one reduction of partial attention states.

\begin{figure*}[t]
	\centering
	\includegraphics[width=0.95\linewidth]{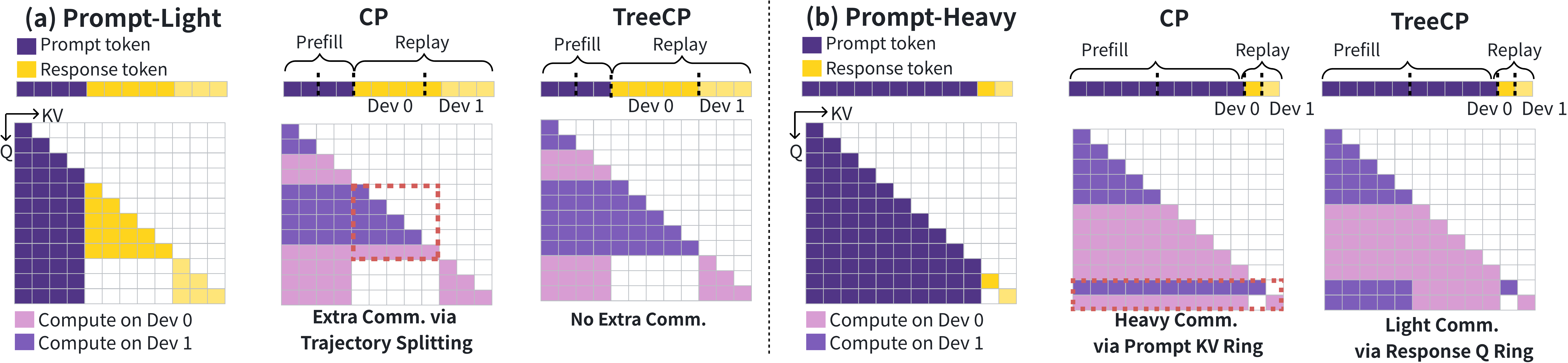}
	\caption{\textbf{When to use TreeMoveKV or MoveQ.}
		\textbf{(a) Prompt-light / response-dominant replay:}
		ordinary CP partitions a flattened sequence and may mix sibling
		response paths. TreeMoveKV retains move-KV execution while placing
		complete response paths or path groups in lane-local rings.
		\textbf{(b) Prompt-heavy replay:}
		ordinary CP repeatedly moves large prompt KV for short responses.
		MoveQ keeps prompt KV stationary and communicates response queries
		and partial reduction states instead.}
	\label{fig:comm}
	\vspace{0em}
\end{figure*}

\noindent\textbf{Communication model.}
Let $\rho=L_P/L_R$ denote the prompt-to-response ratio, where $L_P$ is the prompt length and $L_R$ is the response replay length proxy.
Let $H_q$ and $H_{kv}$ be the query and KV head counts.
Ignoring topology constants, the repeated point-to-point communication of TreeMoveKV and MoveQ is approximated as:
\begin{align}
	C_{\textsc{TreeMoveKV}}
	&\approx
	2(L_P+L_R)H_{kv},
	\label{eq:treemovekv_comm_cost}\\
	C_{\textsc{MoveQ}}
	&\approx
	L_R H_q + T_{\mathrm{red}}(L_R),
	\label{eq:moveq_query_cost}
\end{align}
where $T_{\mathrm{red}}(L_R)$ denotes the calibrated overhead of the one-time reduction over partial softmax states.
The relative communication cost is therefore
\begin{equation}
	\label{eq:treemovekv_moveq_ratio}
	\frac{C_{\textsc{TreeMoveKV}}}{C_{\textsc{MoveQ}}}
	\approx
	\frac{2(L_P+L_R)H_{kv}}{L_R H_q+T_{\mathrm{red}}(L_R)}.
\end{equation}
As $\rho$ grows, repeated KV movement becomes increasingly expensive relative to query movement plus reduction.
This first-order model is an intuitive approximation intended only to expose the main regime switch, not a rigorous cost estimate: it omits byte counts, collective latencies, and topology effects.
It shows that TreeMoveKV is attractive when response-side replay dominates, while MoveQ becomes attractive when prompt KV is large and response queries are short.
The rigorous cost model used for planning, which accounts for CP degree, replay grid, placement, padding, memory limits, launch overheads, and measured cluster behavior, is given in Sec.~\ref{sec:online_planner}.

\subsection{Online Strategy Planner}
\label{sec:online_planner}
We next introduce AugTree's online planner. As replay trees emerge only after rollout, static offline CP schedules are infeasible. Before GPU dispatch, the planner selects a memory-feasible plan that balances replay work and minimizes the estimated total time for prompt prefill and response replay.


\noindent\textbf{Plan space and feasibility.}
For each replay tree, the planner evaluates plan candidates $c=(d,\pi,\mathbf{r},X)$ 
where $\mathbf{r}=(b,s)$, $d$ is the CP degree, $\pi$ is the execution primitive, $\mathbf{r}$ is the replay grid, and $X$ is the placement. 
Here $b$ is the number of response-replay rounds and $s$ is the number of replay lanes per round. When $b=0$, AugTree does not split prompt prefill from response replay and instead runs ordinary ring CP, \textsc{RingCP}~\cite{liu2024ringattention}, separately over each $(P,R_i)$ pair using the standard causal mask, forming the fused candidate$
(d,\textsc{RingCP},(0,1),\varnothing)$.
The prompt is thus recomputed per branch, which is exactly why this fallback is selected only when split replay is not worthwhile. 

When $b\ge 1$, $\pi\in\{\textsc{TreeMoveKV},\textsc{MoveQ}\}$. 
For \textsc{TreeMoveKV}, the \(d\) ranks form \(s\) disjoint lanes, each with $d_{\mathrm{lane}}=d/s$,
so \(s\) must divide \(d\); each lane runs an independent response-local KV ring. For \textsc{MoveQ}, \(X\) assigns response queries to lanes while prompt-KV shards remain stationary and partial attention states are reduced by query.  The planner keeps only candidates satisfying primitive-specific rank and collective constraints, non-empty active lanes, valid process-group topology, and per-rank memory limits for prompt states, activations, communication buffers, and temporary outputs. It also rejects placements that assign incompatible concurrent roles to the same rank. 
The search is thus exhaustive over the candidate set (degree, primitive, and grid---the decisions profiling shows dominate end-to-end time), while the placement \(X\) per candidate is a balanced makespan assignment, solved exactly by subset DP for $U\le U_{\max}$ (a small constant cap that keeps the assignment bounded) and by largest-first greedy balancing otherwise. It does not span hypergraph cuts, per-layer primitive mixtures, or general DAG schedules.

\noindent\textbf{Replay units and grid assignment.}
For each replay tree, the planner observes the prompt length \( |P| \), response lengths \(\{|R_i|\}_{i=1}^{G}\), group size \(G\), and candidate CP degrees \(\mathcal{D}\). It also uses cluster profiles of compute throughput, point-to-point and collective bandwidth, launch latency, and per-rank memory capacity to estimate execution time and feasibility. For planning, the implementation uses $L_R^{\max}=\max_i |R_i|$ as a conservative replay-length proxy and records the mean response length for diagnostics.
AugTree converts the \(G\) response paths into a bounded set of replay units
$\mathcal{U}=\{u_k\}_{k=1}^{U}$,
where $U\le U_{\max}$, \(U_{\max}\) is an upper bound on the number of replay units, and each unit is one complete response path.
Each unit receives a work estimate \(w_k\) based on its query length, valid Q--KV tile count, and primitive-specific profile.

For a fixed candidate \((d,\pi,b,s)\), the planner assigns each unit to a replay round and lane, 
$X_k=(\tau_k,j_k)$, where $\tau_k\in\{1,\ldots,b\}$ and $j_k\in\{1,\ldots,s\}.$ 
Lanes within a round execute concurrently, while rounds execute sequentially. The assignment minimizes the predicted replay critical path:
\begin{equation}
	\label{eq:lane-assignment}
	\min_X
	\sum_{\tau=1}^{b}
	\max_{j\in\{1,\ldots,s\}}
	\sum_{u_k:X_k=(\tau,j)} w_k,
\end{equation}
subject to single-assignment, per-lane memory, and primitive-specific ownership and topology constraints. AugTree solves this problem with \textsc{AssignGrid}, a bounded dynamic program tracking assigned units, lane loads, and replay rounds. Since \(U\) is capped by the small constant \(U_{\max}\), planning remains fast enough for online execution before GPU dispatch.

\begin{algorithm}[t]
	\caption{AugTree Online Planner}
	\label{alg:treecp-planner}
	\begin{algorithmic}[1]
		\Require Tree $\mathcal{T}=(P,\{R_i\}_{i=1}^{G})$, candidate degrees $\mathcal{D}$
		\State Build bounded replay units $\mathcal{U}$ and weights $\{w_i\}_{i=1}^{U}$
		\State $c^\star \gets \varnothing$, $\widehat{T}^\star \gets \infty$
		\For{$d\in\mathcal{D}$}
		\State $\mathcal{C}_d \gets \{(\textsc{RingCP},0,1)\}\cup \Call{FeasibleSplits}{d}$
		\For{$(\pi,b,s)\in\mathcal{C}_d$}
		\State $X \gets \varnothing$ if $b=0$, else $\Call{AssignGrid}{\mathcal{U},\{w_i\},b,s}$
		\State $c \gets (d,\pi,(b,s),X)$, compute $\widehat{T}(c)$ by Eq.~\ref{eq:planner_score}
		\If{$\widehat{T}(c)<\widehat{T}^\star$}
		\State $c^\star \gets c$, $\widehat{T}^\star \gets \widehat{T}(c)$
		\EndIf
		\EndFor
		\EndFor
		\State \Return $c^\star$
	\end{algorithmic}
\end{algorithm}

\noindent\textbf{Cost model.}
Each feasible split candidate with $b\ge 1$ is scored by estimated prefill time plus replay time:
\begin{equation}
	\label{eq:planner_score}
	\widehat{T}(c)
	=
	\gamma_c\left(
	\widehat{T}_{\mathrm{prefill}}(c)
	+
	\widehat{T}_{\mathrm{replay}}(c)
	\right),
\end{equation}
where $\gamma_c$ is a calibration factor measured from held-out profiling runs.
For $b=0$, the planner scores a standard ring CP pass over the prompt and responses together.
This candidate is selected when tree-specific replay is not worthwhile.

For split candidates, the prefill term constructs the shared prompt KV once:
\begin{equation}
	\label{eq:prefill_cost}
	\widehat{T}_{\mathrm{prefill}}
	=
	\alpha_p \frac{L_P}{d}
	+
	\beta_p L_P\, f_{\mathrm{ring}}(d)
	+
	L_p(d).
\end{equation}
The $\alpha$ term models compute, the $\beta$ term models communication, and $L_p$ models fixed launch/collective overhead during prefill.
The replay term depends on the selected primitive.
For TreeMoveKV, replay keeps move-KV semantics:
\begin{equation}
	\label{eq:treemovekv_replay_cost}
	\widehat{T}_{\mathrm{replay}}^{\mathrm{TMKV}}
	=
	\alpha_{\mathrm{TMKV}} W(\mathbf{r},X)
	+
	\beta_{\mathrm{TMKV}} C_{\mathrm{KV}}(\mathbf{r},X)
	+
	L_{\mathrm{TMKV}}(d,\mathbf{r},X),
\end{equation}
where $W(\mathbf{r},X)$ is the critical-path replay work, $C_{\mathrm{KV}}$ is the KV traffic, and $L_{\mathrm{TMKV}}$ captures scheduling, communication startup, and kernel launch overhead.
For MoveQ, replay keeps prompt KV stationary and reduces partial softmax states:
\begin{equation}
	\label{eq:moveq_cost}
	\widehat{T}_{\mathrm{replay}}^{\mathrm{MQ}}
	=
	\alpha_{\mathrm{MQ}} W(\mathbf{r},X)
	+
	\beta_{Q} C_{Q}(\mathbf{r},X)
	+
	T_{\mathrm{r}}(\mathbf{r},X,d)
	+
	L_{\mathrm{MQ}}(d,\mathbf{r},X).
\end{equation}
Here $C_Q$ is the query-transfer traffic, $T_{\mathrm{r}}$ models the collective reduction over partial softmax states, and $L_{\mathrm{MQ}}$ captures MoveQ communication startup and launch overhead.

\noindent\textbf{Solver summary.}
Algorithm~\ref{alg:treecp-planner} summarizes the planner.
The outer loop enumerates the $b=0$ ordinary ring CP candidate and feasible split candidates $(d,\pi,\mathbf{r})$.
Each candidate is filtered by memory, lane-population, and topology constraints.
For split candidates, \textsc{AssignGrid} computes the placement $X$ with the assignment solver described above, and the complete plan is scored with Eq.~\ref{eq:planner_score}.
MoveQ is not chosen by a hard threshold; Eq.~\ref{eq:treemovekv_moveq_ratio} only informs the communication estimate, while the final decision uses the calibrated end-to-end score.
The dominant planning cost is the candidate enumeration
$O(|\mathcal{D}|K)$,
where $\mathcal{D}$ is the set of candidate CP degrees and $K$ is the number of valid $(\pi,\mathbf{r})$ choices per CP degree; since $U$ unit counts never exceed the constant cap $U_{\max}$, planning overhead grows mildly with response count.

 \begin{figure*}[t]
	\centering
	\includegraphics[width=0.96\textwidth]{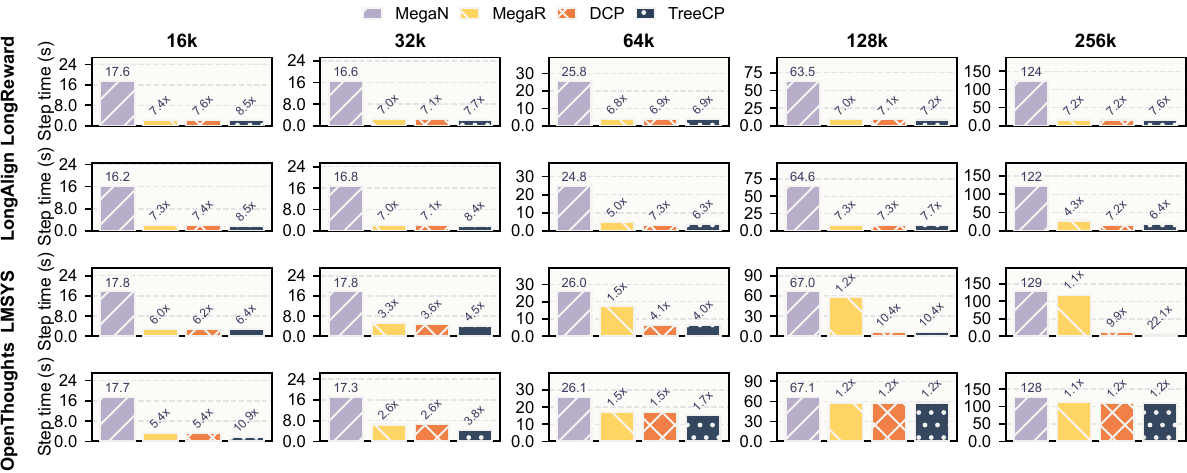}
	\caption{\textbf{End-to-end training-stage step time.}
		Qwen2.5-7B, $G{=}8$, four real post-training workloads, and 16k--256k tokens per step.}
	\label{fig:main}
	\vspace{0em}
\end{figure*}

\subsection{Plan-Driven Execution Engine}
\label{sec:execution_engine}
The engine realizes the selected plan $c=(d,\pi,\mathbf{r},X)$ inside the RLHF trainer, translating it into process groups, token ownership, communication operators, and attention kernels.

\noindent\textbf{Overlapped plan dispatch.}
Planning runs on CPU from rollout lengths while GPUs execute previously planned trees; the small plan is broadcast once at dispatch, keeping planning off the steady-state critical path.

\noindent\textbf{Parallelism-aware group materialization.}
AugTree builds CP groups within each pipeline stage by reusing data-parallel replicas as sequence-parallel ranks; ranks in one PP stage jointly execute one replay tree under degree $d$, while PP preserves normal ordering and DP replicas either run different trees or share gradients. Tensor parallelism is orthogonal: TP groups shard model tensors along the original dimension while sharing the same plan, and all layers of one tree follow one plan.

\noindent\textbf{Adaptive fallback and split replay.}
For $b=0$, the engine runs the whole tree with ordinary ring CP. For split plans ($b\ge 1$), it prefills the prompt once per PP stage and stores per-layer prompt KV as \texttt{past\_key\_value}; prefill always uses the prompt-KV ring path, and the primitive choice applies only to replay. Response branches then replay across the selected rounds and lanes, each call feeding response tokens with cached prompt KV, so losses and gradients are computed without re-running the prompt per branch.

\noindent\textbf{Autograd-compatible attention dispatch.}
AugTree implements TreeMoveKV and MoveQ as drop-in self-attention replacements: the wrapper reads the plan and dispatches each layer to the ring, tree-local KV ring, or Q-moving path. Communication and reduction steps are autograd operations, so backward returns gradients to the original tensor owners---these are the only model-operator changes; losses, gradient synchronization, and optimizer steps remain those of the base trainer.

\section{Experiments}
\label{sec:eval}

\subsection{Setup}
\label{sec:eval-setup}


We evaluate on an 8-node GPU cluster, each equipped with two 64-core CPUs and eight SIMT accelerators. Each accelerator provides 64\,GB HBM, 1.8\,TB/s memory bandwidth, and 37\,Tflop/s peak FP64 throughput. Intra-node CPU--GPU communication uses PCIe Gen5, while inter-node communication uses native RDMA with four 400\,Gbps ports per node. Jobs run via Slurm and \texttt{torchrun} using Python 3.10, PyTorch 2.4.1, OpenMPI 5.0.3, and GCC 8.5.0.

\noindent\textbf{Models and protocols.}
We evaluate Qwen2.5-Instruct models at 1.5B, 3B, 7B, and 14B scales~\cite{hui2024qwen2}, using Qwen2.5-7B-Instruct by default, and additionally Llama-3.1-8B-Instruct~\cite{grattafiori2024llama} to confirm cross-model generality. The 7B model has 28 layers, hidden size 3584, 28 query heads, 4 KV heads, head dimension 128, and FFN size 18,944. All methods use bfloat16 and identical optimizer settings. Unless noted otherwise, we set the response group size to \(G{=}8\), and later evaluate scalability with \(G\in\{8,16,32,64\}\). All methods share the same pipeline-parallel layout and differ only in CP execution, isolating the effects of prompt reuse, CP placement, and communication semantics. We report layouts as CP\(\times\)PP\(\times\)DP. Long-context experiments use \(8{\times}4{\times}1\) on 32 accelerators for 16k--128k tokens per step and \(16{\times}4{\times}1\) on 64 accelerators for 256k tokens.

\noindent\textbf{Baselines.}
We compare AugTree with three baselines. \textbf{MegaN} uses Megatron ring CP~\cite{shoeybi2019megatron,liuringattention} without prompt reuse, flattening the \(G\) prompt--response pairs into independent samples. \textbf{MegaR} adds prompt reuse and applies zigzag ring placement~\cite{gu2024loongtrain,zhu2024ring} during response replay. \textbf{DCP}~\cite{jiang2025dcp}, a dynamic CP, reuses the prompt and dynamically partitions replay blocks using a graph objective, while retaining KV-rotation semantics. All methods use identical data and model configurations; AugTree additionally exploits the prompt--response tree and adaptively selects \textsc{TreeMoveKV} or \textsc{MoveQ}.

%
%
%

\noindent\textbf{Workloads.} 
Following DCP~\cite{jiang2025dcp}, replay workloads are constructed from LongReward~\cite{zhang2025longreward}, LongAlign~\cite{bai2024longalign}, LMSYS-Chat-1M~\cite{zheng2024lmsys}, and OpenThoughts~\cite{guha2025openthoughts}.
Each corpus is converted into prompt--response pairs using deterministic rules independent of the evaluated CP method.  For LongReward, the prompt combines the long context and query, while the response is the provided or preferred answer. For LongAlign, the user message forms the prompt and the assistant message the response. Each LMSYS user--assistant turn is treated as one sample, whereas OpenThoughts uses the problem and distilled reasoning trace as the prompt and response. These corpora span prompt-heavy alignment, short conversations, and response-heavy reasoning.

For the main sweep, we set \(G{=}8\), and use per-step token budgets $
G(L_P{+}L_R)\!\in\!\{16\text{k},32\text{k},64\text{k},128\text{k},256\text{k}\}.$ 
Given budget \(B\), the target per-branch length is \(B/G\). We tokenize valid pairs with the Qwen tokenizer,  and  then rank them by $
\left|\log(B/(G(L_P+L_R))\right|,$ 
breaking ties by corpus order. This controls the training-token budget while preserving each corpus's prompt--response regime. For group-size scalability, we vary \(G\) on the 32k-token LongAlign workload while keeping total tokens per step fixed.

\noindent\textbf{Metrics.}
Our primary metric is end-to-end \emph{training-stage} step time, averaged over steady-state iterations.
This measures the update stage of post-training: replaying scored trajectories and updating the policy.
We also report attention forward/backward time, prompt prefill time, response replay time,
context-parallel communication volume, planner solve time, and step time under model-size and response-group-size scaling.
Our pipeline-share profiling additionally measures the update fraction of the full RL pipeline.

\begin{figure*}[t]
	\centering
	\includegraphics[width=0.96\textwidth]{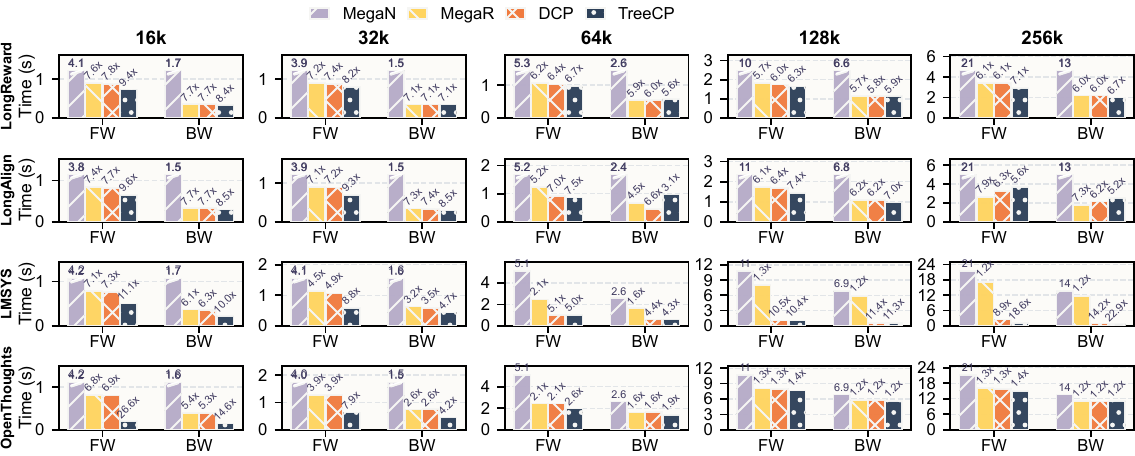}
	\caption{\textbf{Attention forward/backward time.}
		Qwen2.5-7B, $G{=}8$, four real post-training workloads, and 16k--256k tokens per step.}
	\label{fig:attention-breakdown}
	\vspace{0em}
\end{figure*}

\subsection{Main Results}
\label{sec:eval-main}

\noindent\textbf{End-to-end step time.}
Fig.~\ref{fig:main} reports training-stage step time on four real prompt--response distributions:
\textit{LongReward}, \textit{LongAlign}, \textit{LMSYS}, and \textit{OpenThoughts}.
Each column fixes the per-step token budget $G(L_P{+}L_R)$ from 16k to 256k, and each panel compares the four methods.

Across the 20 dataset--length points, AugTree is on average 7.08$\times$ faster than MegaN,
2.63$\times$ faster than MegaR, and 1.18$\times$ faster than DCP.
The largest gains appear on LMSYS at 256k, where short responses make repeated prompt processing
and prompt-KV movement especially expensive.
AugTree loses to DCP on only 4 of the 20 points: two LMSYS cases differ by at most 1.2\%, while
LongAlign at 64k/256k favors DCP because its graph-balanced placement happens to fit those sampled layouts.
Across all 20 AugTree plans, the planner selects MoveQ in 9 cases and move-KV plans in 11 cases,
including 8 TreeMoveKV plans and 3 fallback ring plans.
This shows that both communication semantics are needed across realistic prompt--response regimes.

These gains come from exploiting both sources of structure in post-training replay:
prompt reuse explains the large gap over MegaN, while the gains over MegaR and DCP show that
reuse alone is insufficient.

\begin{figure}[t]
	\centering
	\includegraphics[width=\linewidth]{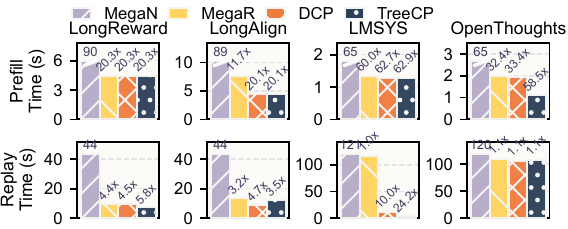}
	\caption{\textbf{Prompt prefill and response replay time.}
		Qwen2.5-7B, $G{=}8$, four real post-training workloads, and 256k tokens per step.}
	\label{fig:prefill-replay}
	\vspace{0em}
\end{figure}

\subsection{Component Ablations and Breakdown}
\label{sec:eval-breakdown}

\noindent\textbf{Attention forward/backward breakdown.}
Fig.~\ref{fig:attention-breakdown} reports per-accelerator average attention forward and backward time over the
same grid as Fig.~\ref{fig:main} (counters divided by 32 for 16k--128k, 64 for 256k).
By attention time, AugTree is on average 7.94$\times$ faster than
MegaN, 2.52$\times$ faster than MegaR, and 1.31$\times$ faster than DCP, confirming that the end-to-end
gains mainly come from reducing replay attention overhead.
Low-gain cases mostly occur on LongAlign, where
DCP's graph-balanced KV rotation is already competitive.

\noindent\textbf{Prompt prefill versus response replay.}
Fig.~\ref{fig:prefill-replay} decomposes the 256k-token cases into prompt prefill and response replay.
MegaN spends substantially more time in prefill because it rebuilds the shared prompt for
each response path, while AugTree, MegaR, and DCP compute it once.

Replay is the main differentiator after prompt reuse. At 256k, AugTree reduces replay time by
8.67$\times$ over MegaN, 6.70$\times$ over MegaR, and 1.37$\times$ over DCP on average,
showing that AugTree improves the replay path itself via branch locality and the cheaper primitive,
beyond removing duplicate prompt computation.

\begin{figure}[t]
	\centering
	\includegraphics[width=\linewidth]{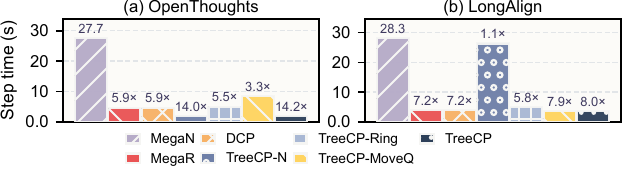}
	\caption{\textbf{Step-time ablations.}
		Qwen2.5-7B, $G{=}8$, g16k per group, 8 accelerators. AugTree(auto) selects the primitive per group.}
	\label{fig:ablation-step-time}
	\vspace{0em}
\end{figure}

\noindent\textbf{Step-time ablations.}
The four execution strategies form a natural component ablation: MegaN uses flattened ring CP
without prompt reuse; MegaR adds prompt reuse but still uses linear ring placement; DCP further
adds dynamic block placement but keeps move-KV execution; AugTree adds prompt--response tree
awareness and primitive selection. Hence MegaN $\rightarrow$ MegaR isolates the benefit of prompt
reuse, MegaR $\rightarrow$ DCP isolates dynamic move-KV placement, and DCP$\rightarrow$AugTree
isolates tree-aware replay with primitive selection. To validate these components directly,
we re-run the full ladder end-to-end on a single 8-accelerator node (no PP/DP) for the g16k
cases, also including forced-primitive variants AugTree-Ring, AugTree-MoveQ, and AugTree-N
(Fig.~\ref{fig:ablation-step-time}). The results confirm the decomposition: prompt reuse
(MegaN $\rightarrow$ MegaR) removes most of MegaN's cost, the planner's automatic selection
(MegaR$\rightarrow$AugTree) gives a further 1.1--2.4$\times$ gain, and forcing the wrong
primitive (e.g., AugTree-MoveQ on response-heavy OpenThoughts, 8.47\,s vs.\ 1.99\,s for
AugTree(auto)) erases the benefit---each design decision contributes, and none is dispensable.

\noindent\textbf{Update share of the full RL pipeline.}
Profiling rollout (8 accelerators, Qwen2.5-7B generation), reward scoring
(2 accelerators, Qwen3-3B classifier), and update (the 8-accelerator Megatron
configuration above) separately at the same per-step tree budget
(\(G{=}8\), 16k tokens), the update share of the pipeline depends mainly on the
prompt--response ratio, ranging from 73\% on the prompt-heavy LongAlign
distribution down to 17\% on the response-heavy OpenThoughts distribution,
where 1907-token responses push rollout to 83\% of pipeline time.
Consequently, AugTree's update-side speedup (up to \(14\times\) over MegaN
and \(2.4\times\) over DCP) yields the largest end-to-end pipeline gains on
prompt-heavy workloads---up to \(2.7\times\) for the full
rollout\(+\)reward\(+\)update pipeline: AugTree is designed for the update
stage, yet on prompt-heavy distributions it benefits the entire pipeline.

\subsection{Communication and Planner Analysis}
\label{sec:eval-comm-planner}

\begin{figure}[t]
	\centering
	\includegraphics[width=0.9\linewidth]{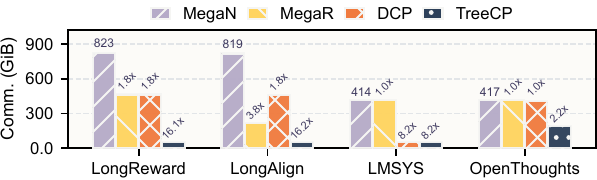}
	\caption{\textbf{CP attention communication volume.}
		Qwen2.5-7B, $G{=}8$, four real post-training workloads, and 256k tokens per step.}
	\label{fig:cp-comm}
	\vspace{0em}
\end{figure}

\begin{figure}[t]
	\centering
	\includegraphics[width=0.9\linewidth]{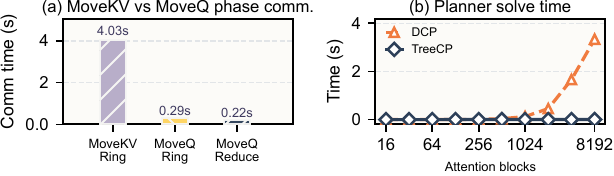}
	\caption{\textbf{MoveQ phase costs and planner solve time.}
		Qwen2.5-7B on 8 accelerators, $G{=}8$.
		(a) MoveKV ring versus MoveQ phase communication time on g16k LongAlign, CP=8;
		(b) planner solve time, 512-token blocks, graph sizes 16--8{,}192.}
	\label{fig:moveq-phase}
	\vspace{0em}
\end{figure}

\noindent\textbf{CP communication.}
Fig.~\ref{fig:cp-comm} reports CP ring-stage attention communication bytes for the 256k workloads.
This metric isolates repeated ring communication in the attention path.
At 256k, AugTree reduces ring-stage communication by avoiding sibling-response traffic and by replacing
repeated prompt-KV movement with query-side movement when prompts dominate.
On the two prompt-heavy cases where AugTree selects MoveQ, AugTree reduces profiled ring-stage
communication by 16.12$\times$, 6.68$\times$, and 9.06$\times$ on average over MegaN, MegaR, and DCP,
respectively.
On response-heavy cases where AugTree selects TreeMoveKV, it reduces profiled ring-stage communication
by 5.15$\times$, 5.14$\times$, and 1.56$\times$ over the same baselines.
This result matches the design: MoveQ is useful when large shared prompt KV would otherwise be
moved repeatedly, while TreeMoveKV is useful when response-side replay dominates and sibling-response
traffic should be avoided. This metric excludes MoveQ's final collective reduction; its small latency impact is quantified in Fig.~\ref{fig:moveq-phase}(a).

\noindent\textbf{MoveQ phase costs.}
Fig.~\ref{fig:moveq-phase}(a) resolves the two MoveQ phases directly:
the query-side ring rotation and the softmax-state reduction together take about one eighth of the
move-KV ring time they replace (0.29+0.22\,s vs. 4.03\,s per step), confirming that swapping
the primitive for prompt-heavy groups is where much of AugTree's communication gain comes from.

\noindent\textbf{Planner overhead.}
In Fig.~\ref{fig:moveq-phase}(b), solve time grows with the replay graph from 16 to 8{,}192 attention
blocks ($G{=}8$, 512-token blocks), with the candidate CP degree scaling up to 128.
DCP constructs and partitions a block-level dependency hypergraph for each candidate degree, causing
solve time to grow from 0.17\,ms at 16 blocks to 3336.5\,ms at 8{,}192 blocks.
AugTree instead enumerates bounded tree-aware candidates over $(d,\pi,s,X)$ and scores them from
precomputed block statistics; its solve time grows only from 0.018\,ms to 1.70\,ms over the same range,
three orders of magnitude faster at the largest size.
Newly generated trees can thus be planned after rollout and before GPU dispatch.

\subsection{Scalability and Correctness}
\label{sec:eval-scale-correctness}

\begin{figure}[t]
	\centering
	\includegraphics[width=\linewidth]{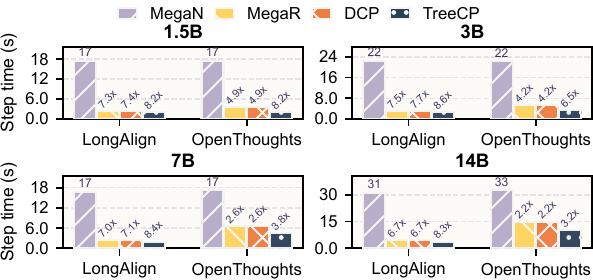}
	\caption{\textbf{End-to-end step time across model sizes.}
		Qwen2.5-Instruct models from 1.5B to 14B on 32k-token LongAlign and OpenThoughts workloads.}
	\label{fig:scale-model}
	\vspace{0em}
\end{figure}

\noindent\textbf{Model scalability.}
Fig.~\ref{fig:scale-model} evaluates the same execution strategies across Qwen2.5 model sizes on
32k-token LongAlign and OpenThoughts workloads.
Across the eight model--dataset points, AugTree is on average 6.99$\times$ faster than MegaN,
1.37$\times$ faster than MegaR, and 1.36$\times$ faster than DCP.
For both LongAlign and OpenThoughts, MegaN scales poorly because it repeats prompt computation
for each response path. MegaR and DCP reduce this duplication, but AugTree still improves over them
because replay placement and communication primitive selection remain important as model compute grows.
The relative gain is smaller on 14B OpenThoughts, where dense model computation becomes a larger
fraction of the step time and therefore reduces the fraction exposed to CP optimization.
AugTree requires no architecture-specific changes: for standard MHA/GQA/MQA attention the semantics are unchanged, and Eq.~(4) explicitly captures the query/KV head counts, which shift the TreeMoveKV--MoveQ crossover. MLA changes only the KV representation and MoE only adds expert-parallel traffic, potentially shifting the optimal plan but not the underlying shared-prefix dependency; as in recent CP systems~\cite{jiang2025dcp,wang2025flexsp}, we evaluate a single family and show gains persist from 1.5B to 14B.

\begin{figure}[t]
	\centering
	\includegraphics[width=\linewidth]{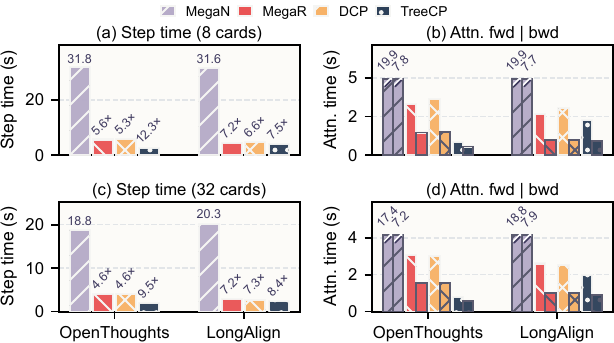}
	\caption{\textbf{Cross-model generality.}
		Llama-3.1-8B, $G{=}8$, g16k per group; (a,b) 8 accelerators on one node,
		(c,d) 32 accelerators across four nodes (same CP$\times$PP$\times$DP$=$8$\times$4$\times$1
		logical layout); step time in (a,c), attention fwd/bwd time in (b,d).}
	\label{fig:llama-step-time}
	\vspace{0em}
\end{figure}

\noindent\textbf{Cross-model family.}
Fig.~\ref{fig:llama-step-time} repeats the g16k comparison on Llama-3.1-8B, at 8 accelerators
on one node (a,b) and 32 accelerators across four nodes (c,d).
At 8 accelerators, AugTree is 12.3$\times$ and 7.5$\times$ faster than MegaN on OpenThoughts and
LongAlign, and 2.2$\times$ and 1.1$\times$ faster than MegaR, confirming that the plan shape, not
the Qwen tokenizer or attention layout, drives the gains. The LongAlign attention panel shows the
smallest AugTree margin there because its prompt-heavy layouts also favor DCP-style placement, as
in Fig.~\ref{fig:attention-breakdown}. The gains persist on the four-node 32-accelerator
configuration: AugTree is 9.5$\times$ and 8.4$\times$ faster than MegaN on OpenThoughts and
LongAlign, and 2.3$\times$ and 1.2$\times$ faster than MegaR, so the benefit also carries over
the node boundary where replay traverses inter-node CP rings.

\begin{figure}[t]
	\centering
	\includegraphics[width=\linewidth]{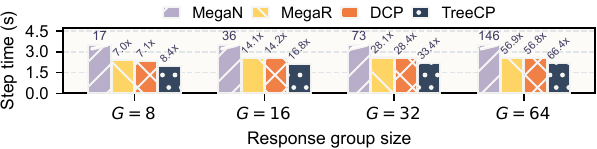}
	\caption{\textbf{End-to-end step time across response group sizes.}
		Qwen2.5-7B, 32k-token LongAlign workload, and $G$ from 8 to 64.}
	\label{fig:group-size-sweep}
	\vspace{0em}
\end{figure}

\noindent\textbf{Group-size scalability.}
Fig.~\ref{fig:group-size-sweep} evaluates response group sizes $G\in\{8,16,32,64\}$ on the
32k-token LongAlign workload while keeping the total training tokens per step fixed.
As $G$ increases, MegaN treats increasingly many prompt--response paths as independent sequences and
repeats prompt computation, while AugTree stays flat (1.99--2.21\,s) by reusing the prompt and selecting
MoveQ for these prompt-heavy groups.
Across the four group sizes, AugTree is on average 31.25$\times$ faster than MegaN, 1.19$\times$
faster than MegaR, and 1.18$\times$ faster than DCP;
MegaR and DCP are close because DCP still selects a move-KV ring plan.

\noindent\textbf{Correctness of AugTree execution.}
AugTree changes the distributed attention schedule but not the attention dependency or RL loss: both
primitives compute the same attention result (Sec.~\ref{sec:primitive_selection}), differing only in
tensor ownership and communication direction, and their communication and reduction steps are autograd
operations that return gradients to the original tensor owners. Prompt gradients likewise accumulate on
the shared prompt states before propagating through the original prefill graph (Sec.~\ref{sec:prefill_once}).
Token-level losses, gradient synchronization, and optimizer updates are unchanged. Validation runs confirm
matching loss trajectories against the move-KV baselines under identical replay inputs.

\subsection{Discussion and Limitations}
\label{sec:discussion}

\noindent\textbf{Workload generality.}
AugTree is driven by the attention dependency structure rather than a specific model family or RL loss: one shared causal prefix fans out into independent autoregressive branches---each response attending to the shared prompt and its own prefix---so group-based objectives~\cite{shao2024deepseekmath,yu2025dapo,guo2025deepseek}, which differ mainly in reward normalization, filtering, or loss weighting, all apply without changing the planner or engine, and the same structure directly covers RAG, multi-turn dialogue, and agentic trajectories that share such a root. AugTree does not exploit deeper internal prefix sharing or general DAGs with cross-branch dependencies or reconvergence; if such workloads are replayed as independent trajectories, AugTree still exploits their common root, and general DAG-aware CP is left to future work.

\section{Related Work}
\noindent\textbf{Long-Context Context Parallelism.}
Long-context training systems use context parallelism to split a single long sequence across devices.
RingAttention~\cite{liuringattention} and Ring Flash Attention~\cite{zhu2024ring} shard the sequence dimension and rotate KV blocks.
DeepSpeed Ulysses~\cite{jacobs2023deepspeed} shards attention heads and uses all-to-all tensor redistribution.
USP~\cite{fang2024usp}, LoongTrain~\cite{gu2024loongtrain}, and TransformerEngine~\cite{nvidia2024transformer} compose sequence and head dimensions to scale long-context training.
Recent systems further optimize kernels, packing, load balance, or flexible placement, including FlashMask~\cite{wang2025flashmask}, JENGA~\cite{wang2025jenga}, SAS~\cite{zhou2026sas}, Hierarchical Balance Packing~\cite{yao2026hierachical}, FlexPipe~\cite{zhao2025flexpipe}, FlexSP~\cite{wang2025flexsp}, ByteScale~\cite{ge2025bytescale}, HexiSeq~\cite{liang2026hexiseq}, WLB-LLM~\cite{wang2025wlb}, and NanoCP~\cite{chen2026nanocp}.
These systems target dense or variable linear sequences, local attention work, pipeline balance, heterogeneous placement, or decoding-time scheduling.
AugTree targets a different workload: post-training replay, where one prompt fans out to multiple response paths.
It preserves this tree structure and changes the replay communication object when moving prompt KV is inefficient.
DCP~\cite{jiang2025dcp} is the closest baseline because it partitions block-level attention graphs for dynamic inputs, but it still assumes move-KV execution and treats replay as a generic graph.

\noindent\textbf{RLHF and Post-Training Systems.}
LLM post-training systems optimize the rollout--reward--training pipeline.
GRPO~\cite{shao2024deepseekmath} and DAPO~\cite{yu2025dapo} use group-based reinforcement learning objectives.
OpenRLHF~\cite{hu2024openrlhf} and HybridFlow/veRL~\cite{sheng2025hybridflow} provide RLHF runtimes, while RealHF~\cite{mei2025real}, RLHFuse~\cite{zhong2025optimizing}, AReaL~\cite{fu2026areal}, RhymeRL~\cite{he2025history}, RollPacker~\cite{gao2025rollpacker}, and Seer~\cite{qin2025seer} optimize resource allocation, asynchrony, rollout scheduling, or long-tail latency.
These are orthogonal to AugTree, which optimizes the training update stage and runs inside such runtimes without changing their rollout scheduling or RL objectives.

\section{Conclusion}

We present AugTree, a prompt-aware context-parallel training method for long-context LLM post-training updates: separating prompt prefill from response replay, preserving response-path locality, and selecting TreeMoveKV or MoveQ by communication pattern. A bounded online planner and plan-driven engine make these decisions practical in existing trainers; experiments on real replay workloads show AugTree improves training-stage step time over Megatron ring baselines and DCP. Beyond the move-KV assumption of prior CP, AugTree shows that preserving prompt state and choosing the communication object matter: on prompt-dominant workloads its update-stage speedup also yields the largest full-pipeline gains (up to 2.7$\times$).


\bibliographystyle{ACM-Reference-Format}
\bibliography{ref}


\end{document}